\documentclass[preprint,5p,times,twocolumn]{elsarticle}
\biboptions{sort&compress}
\usepackage{amssymb}
\usepackage{amsmath}
\usepackage[hidelinks]{hyperref}
\makeatletter
\def\ps@pprintTitle{%
  \let\@oddhead\@empty
  \let\@evenhead\@empty
  \def\@oddfoot{\reset@font\hfil\@date}
  \let\@evenfoot\@oddfoot
}
\makeatother

\begin{document}

\begin{frontmatter}

\title{Design and initial results from the ``Junior'' Levitated Dipole Experiment}

\author[openstar]{C.S. Chisholm\corref{cor1}}\ead{craig.chisholm@openstar.nz}
\author[openstar]{T. Berry}
\author[openstar]{D.T. Garnier}
\author[openstar]{R.A. Badcock}
\author[openstar]{G. Bioletti}
\author[openstar]{K. Bouloukakis\fnref{kpres}}
\author[openstar]{E. Brewerton}
\author[openstar]{M.A. Buchanan}
\author[openstar]{P.J. Burt}
\author[openstar]{E.V.W. Chambers}
\author[openstar]{K.B. Chappell}
\author[openstar]{P. Coulson}
\author[openstar]{R.J. Davidson}
\author[openstar]{J.P.M. Ellingham}
\author[openstar]{P. Geursen}
\author[openstar]{K. Hamilton}
\author[openstar]{R. Hu}
\author[openstar]{E. Hunter}
\author[openstar]{J.P. Jones}
\author[openstar]{P. Kusay}
\author[openstar]{Z. Lazi\'{c}}
\author[openstar]{B. Leuw}
\author[openstar]{M. Lynch}
\author[openstar]{R. Mataira}
\author[openstar]{M. McCrohon}
\author[openstar]{L. Meadows}
\author[openstar]{J.R. Morris}
\author[openstar]{R. Nowacki}
\author[openstar]{J.V. Purvis}
\author[openstar]{J.H.P. Rice}
\author[openstar]{M. Rutten}
\author[openstar]{S. Schimanski}
\author[openstar]{A. Sharma}
\author[openstar]{M. Siamaki}
\author[openstar]{A. Simpson}
\author[openstar]{T. Simpson}
\author[openstar]{B. Smith}
\author[openstar]{E. Stiers}
\author[openstar]{E. Swanson-Dobbs}
\author[openstar]{J. Todd}
\author[openstar]{E.O.P. Treacher\fnref{npres}}
\author[openstar]{J.D. Tyler}
\author[openstar]{S. Venuturumilli}
\author[openstar]{H.W. Weijers}
\author[openstar]{T. Wordsworth}
\author[openstar]{N. Zhou}
\affiliation[openstar]{organization={OpenStar Technologies Limited},
                       addressline={20 Glover Street, Ngauranga},
                       city={Wellington},
                       postcode={6035},
                       country={New Zealand}}
\cortext[cor1]{Corresponding author}
\fntext[kpres]{Present address: Proxima Fusion GmbH, Fl\"{o}{\ss}ergasse 2, 81369 M\"{u}nchen, Deutschland}
\fntext[npres]{Present address: The MacDiarmid Institute for Advanced Materials and Nanotechnology, School of Chemical and Physical Sciences, Victoria University of Wellington, PO Box 600, Wellington 6140, New Zealand}

\begin{abstract}
\noindent OpenStar Technologies is a private fusion company exploring the levitated dipole concept for commercial fusion energy production. OpenStar has manufactured a new generation of levitated dipole experiment, called ``Junior", leveraging recent advances made in high-temperature superconducting magnet technologies. Junior houses a $\sim5.6$~T REBCO high-temperature superconducting magnet in a $5.2$~m vacuum chamber, with plasma heating achieved via $< 50$~kW of electron cyclotron resonance heating power. Importantly, this experiment integrates novel high temperature superconductor power supply technology on board the dipole magnet. Recently OpenStar has completed first experimental campaigns with the Junior experiment, achieving first plasmas in late 2024. Experiments conducted with the full levitated system are planned for 2025. This article provides an overview of the main results from these experiments and details improvements planned for future campaigns.
\end{abstract}

\end{frontmatter}

\section{Introduction}

Dipole confined plasmas were first proposed as a fusion concept by Hasegawa in 1987~\cite{Hasegawa1987} after spacecraft observations of strongly peaked plasma pressure and density profiles in magnetospheres due to strong inward particle pinch caused by turbulence~\cite{Melrose1967,Farley1970}. The magnetospheres of Earth and Jupiter are examples of stable plasmas confined by a simple dipole field found in nature~\cite{Gold1959a,Gold1959} and laboratory dipole magnetic confinement dates back to supported ``terrellae'' experiments conducted by Birkeland in the early twentieth century~\cite{Egeland2005}. The Collisionless Terrella Experiment (CTX) is a supported dipole experiment which studied the phase-space evolution of dipole-trapped energetic plasma in the presence of drift-resonant fluctuations in the 1990s through to the 2010s~\cite{Warren1995,Warren1995a,Grierson2009,Roberts2015,Roberts2015a}. The Space Plasma Environment Research Facility is a supported terrella-like experiment for magnetospheric plasma physics studies~\cite{Qingmei2017,He2023}. Supported magnetic dipoles are limited to very low collisionality regimes where pitch angle scattering into the loss cone is slow enough to compare to cross field turbulent transport~\cite{Hansen2001}.

Magnetic dipoles without mechanical supports are required to study transport with higher collisionality and for fusion relevance. Levitated magnetic dipoles have been investigated in the Levitated Dipole Experiment (LDX)~\cite{Garnier2006} and the Ring Trap 1 (RT-1) experiment~\cite{Morikawa2007}. In particular, the LDX experiment was able to show peaked plasma pressure profiles resulting from a turbulent pinch when the dipole coil was levitated but not while it was supported~\cite{Boxer2010}. Peaked density profiles were also observed in RT-1 where they achieved peak local $\beta > 1$ ($\beta$ is the ratio of the plasma and magnetic pressures)~\cite{nishiura:2015}.

The ``Junior'' experiment, built by OpenStar Technologies, is a new generation of dipole confined plasma experiment which builds on the pioneering work of LDX and RT-1. In both experiments, a toroidal superconducting current ring, the ``core magnet'' is levitated in a vacuum chamber by a second electromagnet, the ``top magnet'' which provides an antisymmetric stabilized levitation magnetic field. The LDX core magnet was made from Nb$_3$Sn which is a low temperature superconductor (LTS)~\cite{Zhukovsky2000} and was inductively charged using an additional charging coil~\cite{Zhukovsky2001}. The RT-1 core magnet is made from first generation high temperature superconductor (HTS) BI-2223 and is charged using current leads in the docked position~\cite{Morikawa2007}. The RT-1 magnet current decays by $0.9$~\% after $8$~hours of levitation due to finite HTS joint resistance~\cite{Morikawa2007}.

The Junior experiment is a proof of concept aiming to replicate results of LDX with a second generation HTS core magnet whilst integrating novel HTS power supplies onboard the core magnet to maintain current during levitation. The successful integration of these novel HTS technologies opens a pathway to higher field magnets which enable the production of magnetically confined plasmas with fusion relevant densities and temperatures. Beyond these initial engineering goals, the Junior facility is an attractive platform for the investigation of fundamental plasma physics phenomena including but not limited to multi-scale plasma turbulence and energy cascades~\cite{Grierson2009}, self-organization phenomena~\cite{Kesner2010,Davis2014}, high-$\beta$ ($\beta>1$) plasma stability regimes~\cite{Garnier1999,Garnier2006,Saitoh2011,Saitoh2020}, ``artificial radiation belt'' formation~\cite{Warren1995,Warren1995a}, non-linear Alfv\'{e}n wave dynamics~\cite{Kozlov2006}, wave-particle and wave-wave interactions in magnetospheric geometries~\cite{Kouznetsov2007,Saitoh2024}, the effect of anisotropy on stability and confinement~\cite{Simakov2000}, and energetic particle dynamics with hot electron interchanges modes~\cite{Levitt2005,Garnier2006}. In this article, we present an overview of the ``Junior'' levitated dipole experiment and some initial results from the first plasma campaign, conducted with the magnet mechanically supported.

\begin{figure}
\includegraphics{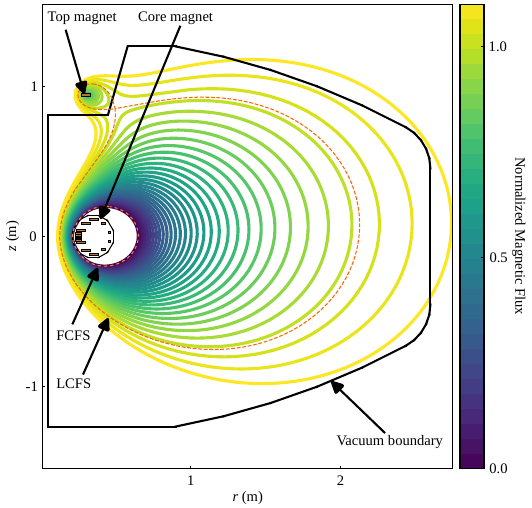}
\caption{\label{fig:Fig1} Radial cross-section showing the magnetic flux of the Junior levitated dipole device.  The floating core magnet is levitated by the fixed top magnet. The solid colored lines show the equilibrium flux contours for a plasma confined by the core magnet operating at $60$~\% of its design current (see Sec.~\ref{sec:coremag}). The FCFS and LCFS are represented by dashed orange lines.}
\end{figure}

\section{Physics Basis}

A levitated dipole is an antisymmetric configuration that confines a torus shaped plasma that surrounds the core magnet. With diamagnetic but no driven currents, the plasma equilibrium consists of a purely poloidal field, {\em i.e.} toroidal confinement without toroidal fields. In Junior, near to the core magnet, there is a first closed flux surface (FCFS) created by the limiting inner bore. The outer boundary or last closed flux surface (LCFS) can be alternatively set by a magnetic separatrix or an outboard limiter. An example of a diverted plasma equilibrium is shown in Fig.~\ref{fig:Fig1}.

In contrast to most magnetic confinement approaches, which require average good curvature and magnetic shear for stability, dipole confined plasmas have bad curvature everywhere outside of the pressure peak and magneto\-hydro\-dynamic (MHD) stability is achieved through plasma compressibility~\cite{Rosenbluth1957,Bernstein1958,Garnier2006a}. Inside the pressure peak, there is absolute good curvature and plasma instabilities should be damped such that transport may be classical as particle drifts conserve poloidal magnetic flux when the toroidal field is zero. For a sufficiently gentle plasma pressure gradient, defined by $\delta(pV^\gamma) \leq 0$, with $p$ the plasma pressure, $V$ the differential flux tube volume, and $\gamma = 5/3$ the ratio of specific heats, dipole confined plasmas are stable to interchange and ballooning instabilities~\cite{Garnier1999}.

The differential flux tube volume is given by $V=\oint\mathrm{d}l/B\propto R^4$ where $R$ is the equatorial radius so that the marginal stability condition, $\delta(pV^{\gamma})=0$, implies a pressure gradient given by $p\propto V^{-\gamma}\propto R^{-20/3}$. This constrains the peak pressure in the dipole according to the edge pressure and flux-tube geometry as $p_{\mathrm{peak}}\leq p_{\mathrm{edge}}(V_{\mathrm{edge}}/V_{\mathrm{peak}})^\gamma$~\cite{Garnier1999}.
This is the stationary profile for low-frequency interchange-like turbulence which leads to the condition $\delta(nV) = 0$. Therefore, the particle density (which is constant within a flux tube) scales as $n=N/V\propto V^{-1}\propto R^{-4}$ where $N$ is the number of particles per flux tube, which is constant. The pressure profile is linked to the temperature and density profiles as $p\propto nT$ leading to a peaked temperature profile given by $T\propto V^{1-\gamma}\propto R^{-8/3}$~\cite{Kesner2010,Boxer2010}.

The location of the pressure peak has been observed to be strongly influenced by the heat source of the plasma~\cite{Boxer2008} but is not strongly influenced by whether the particle source is external or internal~\cite{Garnier2017}. Central heating in a levitated dipole plasma can create a strong central pressure gradient such that $\eta=\mathrm{d}\ln{T}/\mathrm{d}\ln{n}=2/3$~\cite{Kesner2010}. For a marginally stable plasma satisfying $\eta=2/3$, an interchange of flux tubes does not transport net energy and leaves the temperature and density profiles unchanged. Due to adiabatic compression, cool source particles from the edge are heated as they move inwards towards the peak and hot particles from the peak are cooled as they move outwards and expand.

Since the peak plasma pressure depends on the edge pressure, the energy and particle balance is partly determined by the physics of plasma flowing along field lines into the walls. It has been observed that when the core magnet is supported rather than levitated, parallel losses result in density profiles which are approximately uniform in flux space so that the number of particles per flux tube peaks on the outside~\cite{Boxer2010}. 

In Junior, a separatrix may be formed that defines the LCFS when the core magnet has lower current and the top magnet current must be increased to maintain levitation as shown in Fig.~\ref{fig:Fig1}. For some conditions with high plasma $\beta$, the plasma may also shift outward to return to an outboard limited configuration. As the presence of a separatrix may change the stability properties of edge and scrape off layer plasma~\cite{Kesner1997}, this operational flexibility allows the edge stability to be studied experimentally.

\section{Experiment Overview}
The Junior system described in this article was designed and built in under $2$ years at a cost of $<\$10$M~USD and serves as a proof of concept for powering the Core Magnet using an HTS transformer rectifier~\cite{Geng_2025,Rice2024} known as a flux pump~\cite{Klundert1981,Klundert1981a} housed inside the magnet coils. As such, many of the key design choices were based on the design of LDX~\cite{Garnier2006a} which saw electron temperatures and densities of $T_e\sim200$~eV and $n_e\sim10^{18}$~m$^{-3}$~\cite{Kesner2010}.

\begin{figure}
\includegraphics{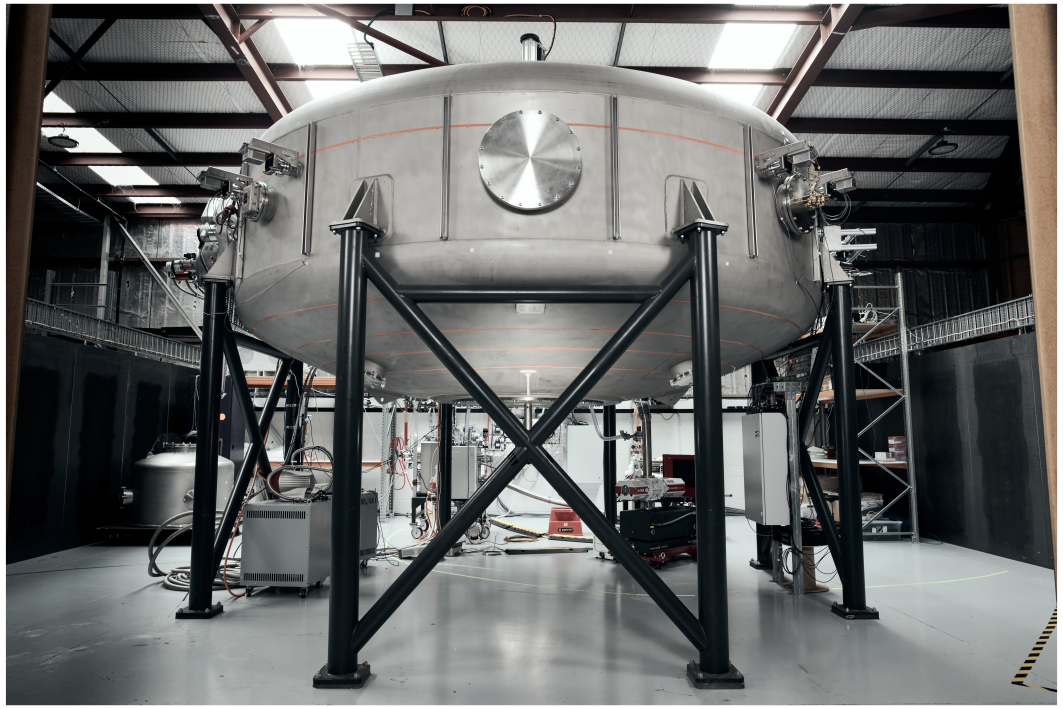}
\caption{\label{fig:Fig2} A photograph of the $5.2$~m wide vacuum vessel for the Junior experiment.}
\end{figure}

\subsection{Vacuum Vessel Size and Material}

The vacuum vessel is made from 304L stainless steel and has an inner radius of $2.6$~m. The maximum internal height of the vacuum vessel is $2.54$~m and the wall thickness is $16.5$~mm. Figure~\ref{fig:Fig2} shows a photo of the vacuum vessel. There are a large number of vacuum ports available for plasma diagnostic installation. Eight ISO-F ports with sizes ranging from DN200 to DN500 are equally spaced around the chamber on the mid-plane. On the bottom of the vessel, four ISO-K flanges (one DN100 and three DN250) are equally spaced on a circle of radius $1.2$~m and each flange is at an angle of $\sim19.5^{\circ}$ from the horizontal. On the top of the vessel, four DN250ISO-K flanges are attached in positions corresponding to the lower ports at an angle of $\sim19.5^{\circ}$ from the horizontal. A custom DN1240 ISO-F flange on the bottom of the vessel allows the core magnet cryostat to be brought in and out of the vessel in one piece, decoupling the magnet and vacuum vessel engineering and allowing for modular upgrades of the core magnet. For recooling, the magnet is docked at the bottom flange and all necessary ground connections are fed through the bottom flange. At the top of the chamber a custom re-entrant flange allows for an independent cryostat for the top magnet.

\begin{table}
\centering
\caption{\label{tab:Tab1} Key parameters of the Junior core magnet.}
\begin{tabular}{lcr}
Parameter & Design Value & This Work\\
\hline
$B_z(0,0)$ & $2.88$~T & $1.20$~T\\
$B_z$ max at windings & $5.63$~T & $2.35$~T\\
$B_R$ max at windings & $2.98$~T & $1.24$~T\\
Current & $1.44$~kA & $600$~A\\
Ampere-turns & $1.5$~MA-turns & $625$~kA-turns\\
Total flux & $0.53$~Wb & $0.22$~Wb\\
Stored energy & $0.55$~MJ & $0.095$~MJ\\
Free bore & $420$~mm & \\
Tape length & $6.3$~km & \\
Inductance & $0.53$~H & \\
Floating mass & $550$~kg & \\
\end{tabular}
\end{table}

\subsection{Core Magnet Description}
\label{sec:coremag}

The Junior core magnet is a $1.44$~kA, $5.7$~T magnet consisting of 14 non-insulated (NI)~\cite{IWASA_2011} solder impregnated HTS coils~\cite{Solder_NI} connected in series. Key design parameters of the core magnet and actual values used for the first plasma campaign are summarized in Tab.~\ref{tab:Tab1}. The magnet is wrapped in multi layer insulation and contained within a 304L stainless steel shell with feed-throughs for cooling and electrical connections while docked. The magnet is nominally cooled to $25$~K using forced helium and all diagnostic sensors are contained within the cryostat shell. The total joint resistance of the coils is estimated to be $8.6 \pm 0.5$~$\mu\Omega$. The vacuum vessel walls and the cryostat shell limit the usable flux to $0.38$~Wb.

\begin{figure}
\includegraphics{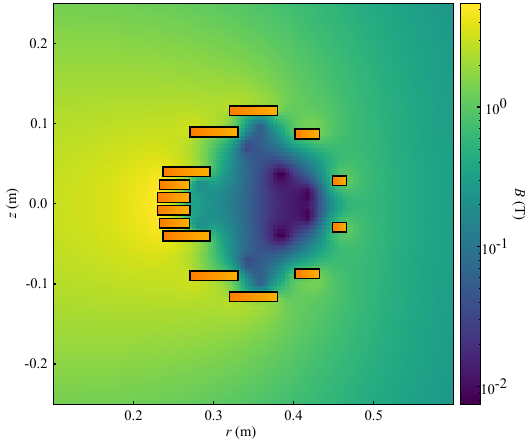}
\caption{\label{fig:Fig3} Cross section view of the Junior core magnet showing the positions of the 14 NI HTS coils (orange boxes) and the calculated magnetic field strength, $B$.}
\end{figure}

During plasma experiments, the core magnet has no connection to external services and current will be maintained using an HTS flux pump (see Sec.~\ref{sec:fluxpump}). The HTS circuit in the flux pump as well as iron yokes used for electromagnet switching are sensitive to magnetic field. A differential evolution algorithm~\cite{Price2005} was used to optimize the distribution of current carrying HTS coils to achieve an adequately low flux such that iron shielding could be used to eliminate stray field at sensitive components. Figure~\ref{fig:Fig3} shows a cross section of the core magnet after coil placement optimization.

\subsection{Flux Pump Description}
\label{sec:fluxpump}

\begin{figure}
\includegraphics{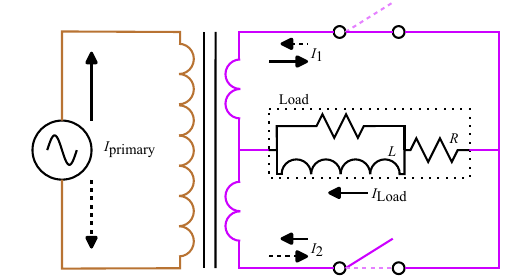}
\caption{\label{fig:Fig4} A schematic representation of a flux pump. The primary circuit (left side) is constructed from normal copper conductor and is driven with a low current oscillating waveform. The secondary circuit (right side) is made of HTS and rectifies the current through the inductive load using electromagnets to drive HTS into the normal conducting state.}
\end{figure}

The core magnet is powered by a superconducting transformer rectifier~\cite{Leuw_2022}, a type of ``flux pump''~\cite{Klundert1981,Klundert1981a}. A schematic representation of this flux pump is shown in Fig.~\ref{fig:Fig4}. An AC waveform is applied on the normal conducting primary circuit. A stepped down AC voltage is produced on the HTS secondary circuit. The waveform is rectified to the load by actuating two switches, which are parallel to the load, out of phase with each other. Each switch consists of lengths of HTS tape kept at an elevated operating temperature. Normal conducting electromagnets are used to apply magnetic field to the switch tape thus driving the switches into a partially resistive state thereby generating voltage. The tape is cut and joined to run back past the electromagnet without capturing magnetic flux in a loop of tape.

The rate of charging of the core magnet is determined by the switch voltages, the inductance of the magnet, and the resistance of the magnet. To maintain current during levitation, the switch voltages must be equal to roughly twice the magnet joint resistance multiplied by the magnet current.

\subsection{Top Magnet Description}

The top magnet is a wax impregnated double pancake HTS coil. The coil has been designed and tested to twice the current required to levitated the core magnet when the core magnet is operating at full current. This means that the core magnet can be levitated when operating at half current, allowing for the formation of a separatrix as shown in fig:~\ref{fig:Fig1}. Key parameters of the top magnet are given in Tab.~\ref{tab:Tab2}.

\begin{table}
\centering
\caption{\label{tab:Tab2} Key parameters of the Junior top magnet.}
\begin{tabular}{lr}
Parameter & Design Value\\
\hline
$B_z(0,0)$ & $0.45$~T\\
$B_z$ max at windings & $2.10$~T\\
Current & $700$~A\\
Ampere-turns & $217$~kA-turns\\
Tape length & $1.2$~km\\
Inductance & $100$~mH
\end{tabular}
\end{table}

\subsection{Plasma Heating Systems}

Plasma heating is via electron cyclotron resonance heating (ECRH). Microwaves with a frequency of $2.45$~GHz are generated by two magnetron heads (Rell GEN15KW2I400-50-0). Each head is capable of generating up to 15 kW of microwave power. The magnetrons are connected to the vacuum chamber using WR430 waveguide and the interface between atmospheric pressure and vacuum pressure is a glass window (Muegge MW0003B-110DC). Similarly to the LDX experiment, we rely on ``cavity heating'' with small first pass absorption and multiple reflections from the vacuum vessel walls to achieve isotropic power distribution~\cite{Hansen2001}.

\begin{figure}[!t]
\includegraphics{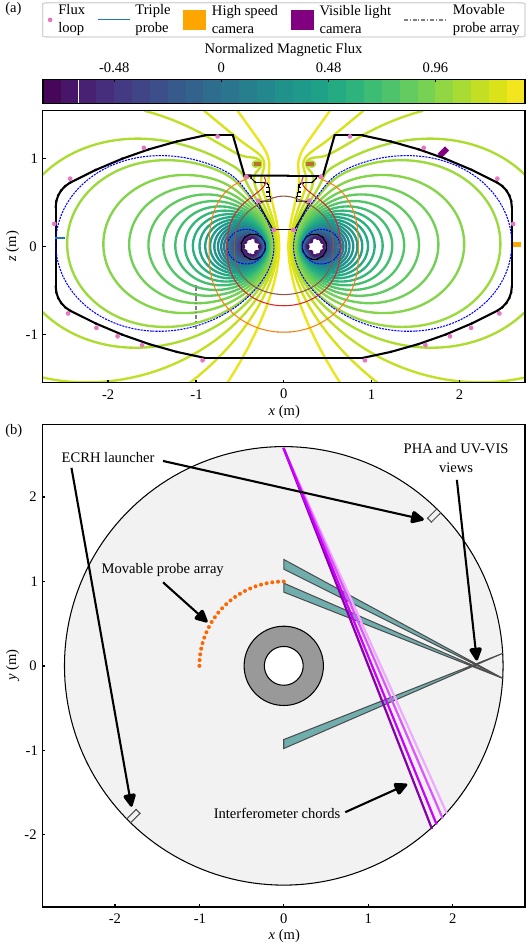}
\caption{\label{fig:Fig5} (a) A side view schematic of the Junior experiment showing the vacuum vessel, the core and top magnets, calculated equilibrium flux contours, and a subset of diagnostics (see main text). The first and last closed flux surfaces are shown as dashed blue lines. ECRH resonance contours at $2.45$~GHz, $6.4$~GHz, and $10.5$~GHz are shown as orange, red, and brown lines respectively. (b) A top view schematic of the Junior experiment showing the vacuum vessel, the core magnet, and a subset of diagnostics (see main text).}
\end{figure}

Microwaves are launched into the vessel using pieces of waveguide cut off at the Vlasov angle. The launchers are positioned at the mid-plane with the H-plane of the waveguide vertical in order to minimize directionality of the launchers and to ensure that microwaves are launched in X-mode~\cite{Stix1992}. Three stub tuners (Muegge GA1006) are installed on the atmosphere side of the waveguide windows to further minimize reflected power. In the near future, two klystron sources will be added to enable multiple frequency ECRH~\cite{Garnier2006,Kesner2010}. These two additional sources are up to $3$~kW at $6.4$~GHz (Varian VA-936R12) and up to $10$~kW at $10.5$~GHz (Varian VA-911 P).

\begin{table}
\centering
\caption{\label{tab:Tab3} Magnetic flux loop parameters.}
\begin{tabular}{llrrr}
Number & Location & $R$ (mm) & $Z$ (mm) & Turns\\
\hline
$1$ & External & $926\pm3$ & $-1280\pm1$ & $4$\\
$2$ & External & $1615\pm4$ & $-1101\pm2$ & $1$\\
$3$ & External & $1895\pm4$ & $-999\pm2$ & $1$\\
$4$ & External & $2135\pm4$ & $-897\pm2$ & $1$\\
$5$ & External & $2458\pm4$ & $-735\pm3$ & $1$\\
$6$ & External & $2616\pm1$ & $261\pm1$ & $1$\\
$7$ & External & $2435\pm4$ & $748\pm3$ & $1$\\
$8$ & External & $1596\pm2$ & $1108\pm1$ & $1$\\
$9$ & Internal & $755\pm2$ & $1259\pm2$ & $5$\\
$10$ & Internal & $427\pm2$ & $800\pm2$ & $10$\\
$11$ & Internal & $292\pm2$ & $525\pm2$ & $20$\\
$12$ & Internal & $110\pm2$ & $193\pm2$ & $35$\\
\end{tabular}
\end{table}

\subsection{Diagnostics}

There is very little restriction on space available for plasma diagnostics since levitated dipole experiments do not have any interlocking coils and importantly, do not need much space for coils on the outer walls of the vacuum vessel. The Junior plasma diagnostic set is illustrated in Fig~\ref{fig:Fig5} along with calculated equilibrium flux contours in Fig~\ref{fig:Fig5}(a). The plasma diagnostics set includes:

\begin{itemize}
    \item A four chord microwave interferometer for line integrated electron density measurements~\cite{Boxer2010}.
    \item a UV-VIS spectrometer for measurement of edge neutral spectra.
    \item Visible light cameras.
    \item Twelve magnetic flux loops (eight external to the vacuum vessel and four internal) for reconstruction of plasma pressure profiles~\cite{Karim2007}.
    \item Two cadmium-zinc-telluride (CZT) x-ray detectors
    \item A sodium-iodide (NaI) x-ray detector.
    \item A silicon drift detector (SDD).
    \item Langmuir probes.
\end{itemize}

The four chord interferometer consists of a single transmitter horn and four receiver horns. As microwaves transmit through the plasma they are phase shifted proportionally to the integrated electron density. This phase shift is measured by beating with a local oscillator which is sent through waveguide on the outside of the chamber and demodulating the signal~\cite{Boxer2010}. The four chords pass through the plasma with tangency radii of $0.93$~m, $0.97$~m, $1.01$~m, and $1.04$~m. We intend to upgrade the interferometer with four additional channels doubling the span of tangency radii. Visible light data is collected using a UV-vis spectrometer (Avantes, Avaspec-2048) and horizontal (Basler a2A2448-105g5cBAS) and top mounted (Basler a2A4096-44g5cBAS) visible light video cameras.

\begin{figure}[t]
\includegraphics{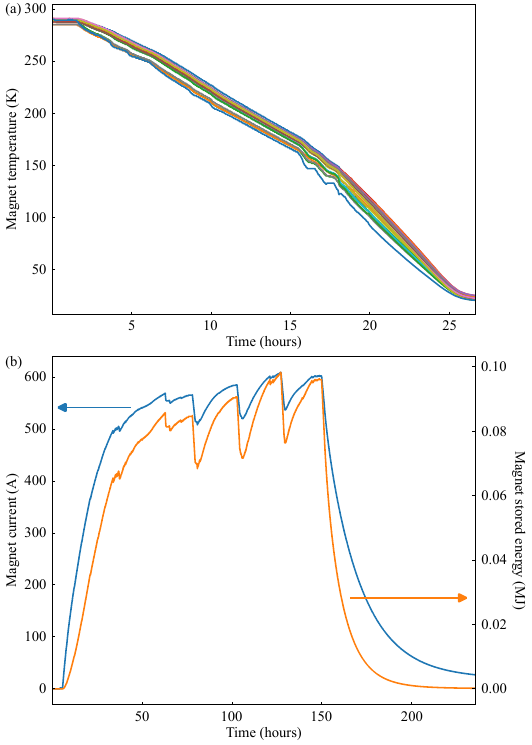}
\caption{\label{fig:Fig6} (a) Cooling of the core magnet to cryogenic temperatures. Cooling rates were maintained at $\sim1$~K/minute with temperature differences around coils controlled at $\lesssim10$~K. (b) Charging the core magnet with a flux pump. Magnet current is shown as a function of time on the left axis and calculated magnet stored energy is shown on the right axis.}
\end{figure}

\begin{figure*}
\includegraphics{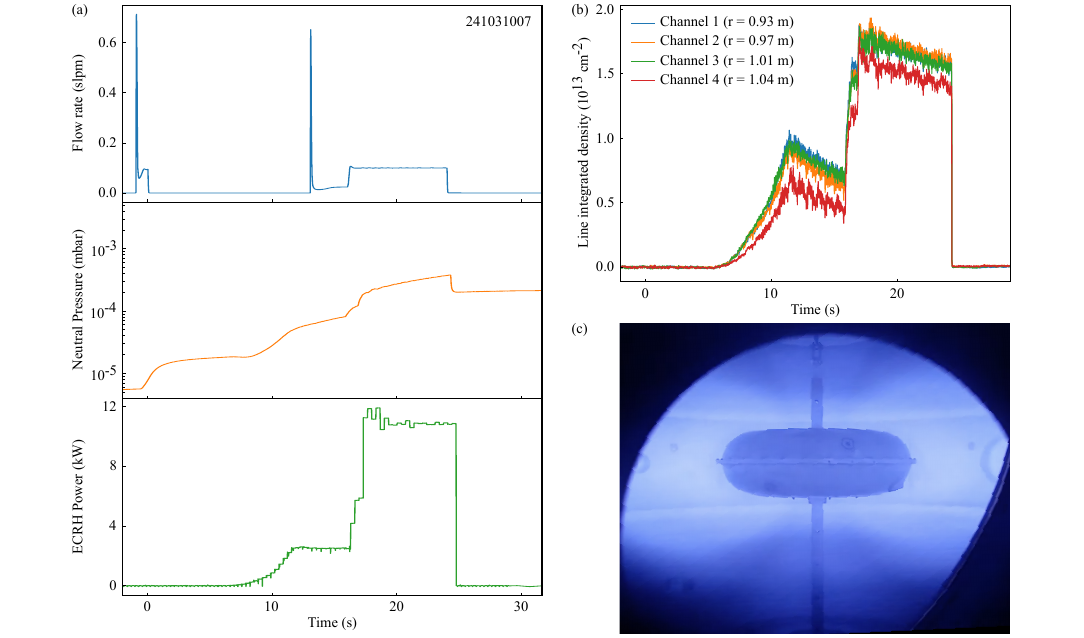}
\caption{\label{fig:Fig7} (a) Gas fueling (top), chamber pressure (middle) and $2.45$~GHz ECRH heating power for plasma discharge $241031007$. (b) Line integrated plasma density showing flat density profiles due to end point losses and modest densities. (c) A color photograph of the plasma.}
\end{figure*}

Magnetic measurements are made using 12 circular flux loops attached to the vacuum chamber which produce a voltage proportional to changing magnetic flux $\varepsilon=-\mathrm{d}\Phi/\mathrm{d}t$. Eight of the flux loops are wound around the outside of the vacuum chamber where their bandwidth is limited to $\sim500$~Hz due to screening currents in the vessel walls~\cite{Karim2007a} but they are able to capture large flux due to their large size. The remaining four flux loops are located inside the chamber and are significantly smaller so have a greater number of turns to compensate for the reduced flux. Two of the internal flux loops are mounted on the top impact attenuator which limits the forces on the core magnet to $\lesssim6g$ in the event of an upward crash ($g$ is the acceleration due to gravity). Table~\ref{tab:Tab3} details the radii and vertical positions of the eight external and four internal flux loops as well as the number of turns in each coil. All of the coils are oriented horizontally and concentric with the core magnet and the positions and radii were optimized for measurement sensitivity given spatial constraints imposed by other mechanical structures. The voltage signals are fed into integrator circuits to obtain magnetic flux which can then be used to reconstruct plasma pressure~\cite{Karim2007,Davis2014}. Additionally, the un-integrated signals of the internal flux loops will be used as a feedback signal in the levitation control loop to compensate for rapid changes in the plasma diamagnetic current which affects the coupling between the top and core magnets.

Since heating in the Junior experiment is achieved using electron cyclotron resonance, the plasma energy is largely electron stored energy with high plasma $\beta$. We use pulse height analyzers to distinguish photon energies from bremsstrahlung radiation and construct histograms from which electron temperature can be inferred. Specifically we employ one NaI detector (Bicron, IA-1378) from which we integrate the counts to obtain total x-ray power between $15$~keV and $3$~MeV with a viewing angle of $22.5^{\circ}$. Two cadmium zinc telluride (CZT) detectors (eV Products SPEAR) with viewing angles of $22.5^{\circ}$ and $27.5^{\circ}$ distinguish photons in the range $10$~keV to $670$~keV with fields of view determined by lead collimators. Finally, a silicon drift detector (AmpTek XR-100SDD) with a viewing angle of $22.5^{\circ}$ measures photons with energy $\lesssim20$~keV and a resolution of $\sim125$~eV. We intend to install an x-ray camera which will allow us to observe the localization of hot electrons and assist with pressure profile reconstruction~\cite{Karim2007}.

Edge density and temperature measurements are made using Langmuir probes. Specifically, we have installed a fixed triple probe~\cite{Polzin2019} and an array of $24$ probes subtending an azimuthal angle of $90^{\circ}$ with equal angular spacing at a radius of $1$~m similar to the probe array which was installed on LDX~\cite{Bergmann2009}. Two thirds of the probes are typically operated in floating potential configuration for measurement of electric field fluctuations~\cite{Garnier2017} and the remaining probes will be operated in ion saturation configuration for local ion density measurements. The vertical probe array can be adjusted over a range of $0.5$~m, as indicated in Fig.~\ref{fig:Fig5}(a). 

\section{Initial Results}

Figure~\ref{fig:Fig6}(a) shows the temperature of the core magnet during a cooling cycle. The magnet was cooled from room temperature to operating temperature ($\sim25$~K) in $\sim24$~hours with a cooling rate of $\sim1$~K/minute and temperature differences around coils controlled at $\lesssim10$~K. A charging campaign with the flux pump is shown in Fig~\ref{fig:Fig6}(b) where the magnet was charged to $\sim600$~A which is $\sim42$~\% of its design current achieving the greatest magnetic stored energy delivered by an HTS flux pump to date. The flux pump was stopped and restarted a number of times to test magnet discharge.

The first plasma campaign on Junior was conducted late 2024 using $^4$He as the fueling gas and a reduced diagnostics set. Seventeen shots were completed over two days of operation. The results in these shots were dominated by neutrals coming off of the wall (additional pumping has since been installed and installation of a glow discharge cleaning system is underway). Because the dipole was not levitated, end point losses resulted in flat line integrated density profiles (also observed in LDX~\cite{Boxer2010}). Figure~\ref{fig:Fig7}(a) shows fueling, pressure, and heating data from a typical plasma shot numbered $241031007$. Line integrated densities from four chord interferometer data are shown in Fig~\ref{fig:Fig7}(b), the flat line integrated density profile indicates a hollow density consistent with a radially localized plasma source. A color photo of the plasma is shown in Fig.~\ref{fig:Fig7}(c).

\section{Conclusion}

The Junior experiment is a new levitated dipole experiment integrating novel HTS power supply technology into a non-insulated HTS magnet. The low cost and rapid construction of the Junior experiment can be attributed to the relative simplicity of levitated dipole as a concept. The same reduced complexity, in particular the lack of interlocking coils, allows a large surface area on the vacuum vessel to be used for prototyping plasma diagnostics. Similarly, large access to the plasma will enable testing of low power ion cyclotron resonance heating in dipole magnetic plasmas as well as initial divertor investigations including strike point control and shaping coils or edge electric field effects from a biased divertor. The first experimental campaign on Junior resulted in 17 plasma shots across two days with the magnet in supported configuration. In this configuration, interferometer data shows a localized plasma density source near ECRH resonance with transport dominated by parallel losses to the stainless steel supports.

The decoupling of magnet and vacuum vessel engineering means that the core magnet can easily be removed and upgraded or swapped out for different designs with relatively little down time. This platform flexibility will enable prototype magnet and fusion technology development in parallel to the construction of larger facilities for fusion relevant plasmas. Homed in New Zealand, we invite external researchers to use the Junior experiment to investigate fundamental plasma physics phenomena of interest to basic plasma science, space physics, and fusion science.

\section*{Acknowledgments}
We thank A. Hasegawa for encouragement and insightful discussions and G.M. Wallace for support and discussions during first plasma operations. We gratefully acknowledge support from the entire OpenStar team. Funding was provided by OpenStar Technologies Limited and we acknowledge additional support from the New Zealand government through Ara Ake Limited.

\bibliography{references.bib}
\end{document}